\documentclass[10pt,conference]{IEEEtran}

\IEEEoverridecommandlockouts
\usepackage{amsmath}
\usepackage{enumitem}
\usepackage{blindtext}
\usepackage{amsfonts}
\usepackage{amssymb}
\usepackage{graphicx}
\usepackage{epstopdf}%
\usepackage[nice]{nicefrac}
\usepackage{svg}
\usepackage{array}
\usepackage{authblk}
\usepackage{tikz}
\usepackage[utf8]{inputenc}
\usepackage{pgfplots} 
\usepackage{pgfgantt}
\usepackage{pdflscape}
\pgfplotsset{compat=newest} 
\pgfplotsset{plot coordinates/math parser=false}
\usepackage{grffile}
\usetikzlibrary{plotmarks}
\usepackage{amsthm}

\usepackage{amsmath}%
\usepackage{MnSymbol}%
\usepackage{wasysym}
\usepackage{subcaption}
\usepackage{soul}

\usepackage[mathscr]{euscript}
\usepackage{cite}
\usepackage{amsfonts}
\usepackage{pgf}
\usepackage{mathtools}
\usepackage{graphicx}
\usepackage[ruled,vlined]{algorithm2e}
\usepackage{array}
\usepackage{tikz}
\usepackage[utf8]{inputenc}
\usepackage{pgfplots} 
\usepackage{pgfgantt}
\usepackage{pdflscape}
\pgfplotsset{compat=newest} 
\pgfplotsset{plot coordinates/math parser=false}
\usepackage{grffile}
\usetikzlibrary{plotmarks}
\usepackage{tikzscale}
\usepackage{caption}
\usepackage{subcaption}

\usepackage{array}
\usepackage{tikz}
\usepackage[utf8]{inputenc}
\usepackage{pgfplots} 
\usepackage{pgfgantt}
\usepackage{pdflscape}
\pgfplotsset{compat=newest} 
\pgfplotsset{plot coordinates/math parser=false}
\usepackage{grffile}
\usetikzlibrary{plotmarks}
\usepackage[T1]{fontenc}
\usepackage{graphicx}
\usepackage{caption}
\usepackage{overpic}
\usepackage{array}
\usepackage{color}
\usepackage{url}
\usepackage{tikz}
\usepackage[utf8]{inputenc}
\usepackage{pgfplots} 
\usepackage{cite}

\usepackage{amsmath}
\usepackage{amssymb}
\renewcommand{\baselinestretch}{1.0}

\usepackage[colorlinks=true,bookmarks=false,citecolor=blue,urlcolor=blue]{hyperref}  %hyperref 
\usepackage[usestackEOL]{stackengine} 

\begin{document}

	\title{Enhancing User Throughput in Multi-panel mmWave Radio Access Networks for Beam-based MU-MIMO Using a DRL Method} 
    % \thanks{Project 6G-LEADER funded by the European Union supported by Smart Networks and Services Joint Undertaking (SNS JU).}

% \vspace{-9mm}
	\author[$\dagger$]{Ramin Hashemi}
	\author[$\dagger$]{Vismika Ranasinghe}
        \author[$\dagger$]{Teemu Veijalainen}
        \author[$\dagger$]{Petteri Kela}
        \author[$\ddagger$]{Risto Wichman}
        % \vspace{-4mm}

	\affil[$\dagger$]{Nokia, Espoo, Finland, \authorcr Emails: {\{ramin.hashemi, vismika.ranasinghe, teemu.veijalainen, petteri.kela\}@nokia.com}}
        \affil[$\ddagger$]{Department of Information and Communications Engineering, Aalto University, Espoo, Finland,\authorcr Email: risto.wichman@aalto.fi}

	\maketitle

	\begin{abstract}
        Millimeter-wave (mmWave) communication systems, particularly those leveraging multi-user multiple-input and multiple-output (MU-MIMO) with hybrid beamforming, face challenges in optimizing user throughput and minimizing latency due to the high complexity of dynamic beam selection and management. This paper introduces a deep reinforcement learning (DRL) approach for enhancing user throughput in multi-panel mmWave radio access networks in a practical network setup. Our DRL-based formulation utilizes an adaptive beam management strategy that models the interaction between the communication agent and its environment as a Markov decision process (MDP), optimizing beam selection based on real-time observations. The proposed framework exploits spatial domain (SD) characteristics by incorporating the cross-correlation between the beams in different antenna panels, the measured reference signal received power (RSRP), and the beam usage statistics to dynamically adjust beamforming decisions. As a result, the spectral efficiency is improved and end-to-end latency is reduced. The numerical results demonstrate an increase in throughput of up to 16\% and a reduction in latency by factors 3-7x compared to baseline (legacy beam management).
	\end{abstract}
	
    \vspace{-3mm}
	\IEEEpeerreviewmaketitle
    \vspace{-3mm}
% 	\IEEEPARstart{T}{he}
    \section{Introduction}
    \vspace{-3mm}
    \bstctlcite{IEEEexample:BSTcontrol}   
    Hybrid beamforming is a technique used in wireless communication systems, particularly in massive multiple-input and multiple-output (MIMO) systems, to efficiently manage the trade-off between performance and hardware complexity with a limited number of radio frequency (RF) chains \cite{Ahmed2018,Junghoon}. It combines both digital and analog beamforming methods. In hybrid beamforming, a reduced number of RF chains are used, which facilitates multi-stream digital baseband processing followed by analog processing to realize antenna beamforming gain. This approach is particularly useful in mmWave communications where the cost and power consumption of fully digital beamforming are prohibitive \cite{Sun2014}. In addition, it allows for energy-efficient operation by reducing the power consumption needed for beamforming so that multiple users are served simultaneously by using different beams.

    The core principle of beam-based multi-user MIMO (MU-MIMO)\footnote{Note that beam-based MU-MIMO framework is enabled by hybrid beamforming architecture for simultaneous transmission to multiple directions via multiple antenna panels.} is to exploit the spatial dimension (SD) of the wireless channel to increase the capacity. By using advanced signal processing algorithms, the system can determine the optimal beamforming vectors that maximize the signal strength at the intended user's location while minimizing interference to other users. This is achieved through the use of channel state information (CSI). Recent research has extensively explored hybrid beamforming and user scheduling for mmWave and massive MIMO systems. Classical approaches such as greedy user scheduling under hardware constraints have been studied in \cite{choi2018user}, while hybrid analog–digital beamforming design for wideband mmWave orthogonal frequency-division multiplexing (OFDM) was developed in \cite{sohrabi2017hybrid}. Deep learning has been increasingly applied to reduce the complexity of channel estimation and hybrid beamforming \cite{elbir2019family,gao2022data}. More recent work has addressed scheduling under imperfect CSI and complexity concerns in OFDM hybrid beamforming \cite{liu2025low}, as well as deep learning–enhanced hybrid beamforming strategies \cite{abir2024deep}. Compared to these studies, the learning-based proportional-fairness scheduling framework in \cite{agheli2025learning} emphasizes a two-timescale protocol and explicitly quantifies the trade-off between fairness and computational complexity. However, using supervised learning techniques involves collecting huge datasets, which is impractical particularly in practical network settings, and generalizing the models to wireless environment conditions. Thus, leveraging reinforcement learning (RL) which learns a policy by interacting with wireless environment is a clear advantage in such scenarios.

    In our previous work in \cite{Hashemi2024}, an RL-assisted beam selection approach studied the analog beamforming use-case for mmWave radio access networks. Single antenna panel gNBs were considered (with only one RF chain), which simplifies the beam selection problem to two dimensions, that is, exploiting beam usage statistics and measured RSRP values for beam selection. However, three dimensions have to be considered in multi-panel gNBs as there will be beams simultaneously activated, which could interfere. In other words, spatial domain characteristics of the grid-of-beams (GoBs) together with aforementioned dimensions add up to the available dimensions for beam selection. Thus, the main novelty of this work lies in proposing a framework for beam management by employing beam usage (or beam popularity), measured RSRP values, and spatial domain cross-correlations in beam-based MU-MIMO technology with multiple antenna panels. Due to the high complexity associated with the proposed throughput optimization problem, an RL approach is used to address the issue.

    % \vspace{-1mm}
    \section{System Model and Problem Formulation}
    \label{SysModelSec}
\begin{figure}[t]
		\centering
		\includegraphics[trim = 1cm 0cm 1cm 0cm,scale=0.55]{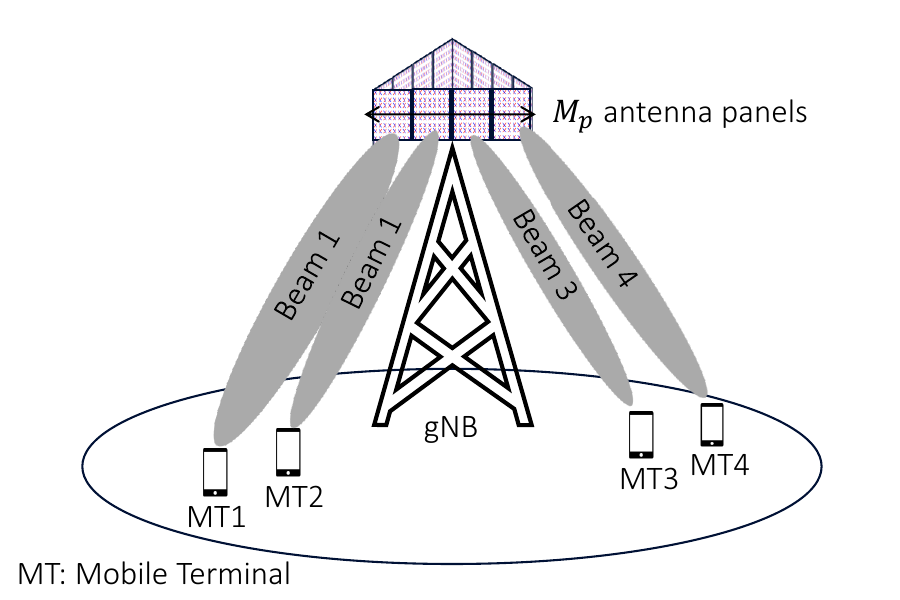}
        % \includesvg[figs/SystemModel.svg]
		\caption{Beam selection approach in beam-based MU-MIMO with $M_p=4$ antenna panels. The activated beams in each panel are spatially separated from beams within other panels, however, spatially closer beams could have a non-zero cross-correlation.}
		\label{fig:0}
    \end{figure}

    We consider the downlink (DL) of a cellular network consisting of $M$ base stations (gNBs) jointly serving $U$ mobile terminals (MTs) where the set of all MTs is denoted as $\mathcal{U}=\{1,2,\cdots,U\}$. Each gNB is equipped with $M_p$ antenna panels as illustrated in Fig.~\ref{fig:0}, where each panel is implemented as a uniform squared planar array (USPA). For directional transmission, every panel supports predefined $L_p$ analog beams based on a GoB structure. Therefore, the total number of beams across all antenna panels is $B=M_p \times L_p$. For simplicity, we collect all beams across all panels into the set $\mathcal{B} = \{1,2,\dots,B\}$, where each beam $b \in \mathcal{B}$ can be indexed by a panel $p$ and a local index $j \in \{1,2,\dots,L_p\}$ such that $b = pL_p + j$. The system follows an analog beamforming architecture with a single RF chain per panel. This implies that only one beam from each panel can be activated within a transmission time interval (TTI). Accordingly, there will be maximum $M_p$ beams activated simultaneously. The MTs are uniformly distributed over the service area and follow mobility profiles with predefined velocities. The wireless propagation environment is modeled using the standardized 3GPP three-dimensional spatial channel model \cite{3GPPTR38.901}. All essential 5G new radio (NR) functionalities, including physical (PHY) layer procedures, medium access control (MAC), and beam management mechanisms, are included. In compliance with 3GPP specifications, each gNB performs beam sweeping across all candidate beams from every panel during a beam-switching cycle of duration $T_{\mathrm{bs}}$ (e.g., $40$ ms). MTs measure the RSRP of the beams, and a subset of the measurements is fed back to the serving gNB, which determines the active beams \cite{Abinader2021}.  
    
\begin{figure}[t]
		\centering
		\includegraphics[width=\linewidth]{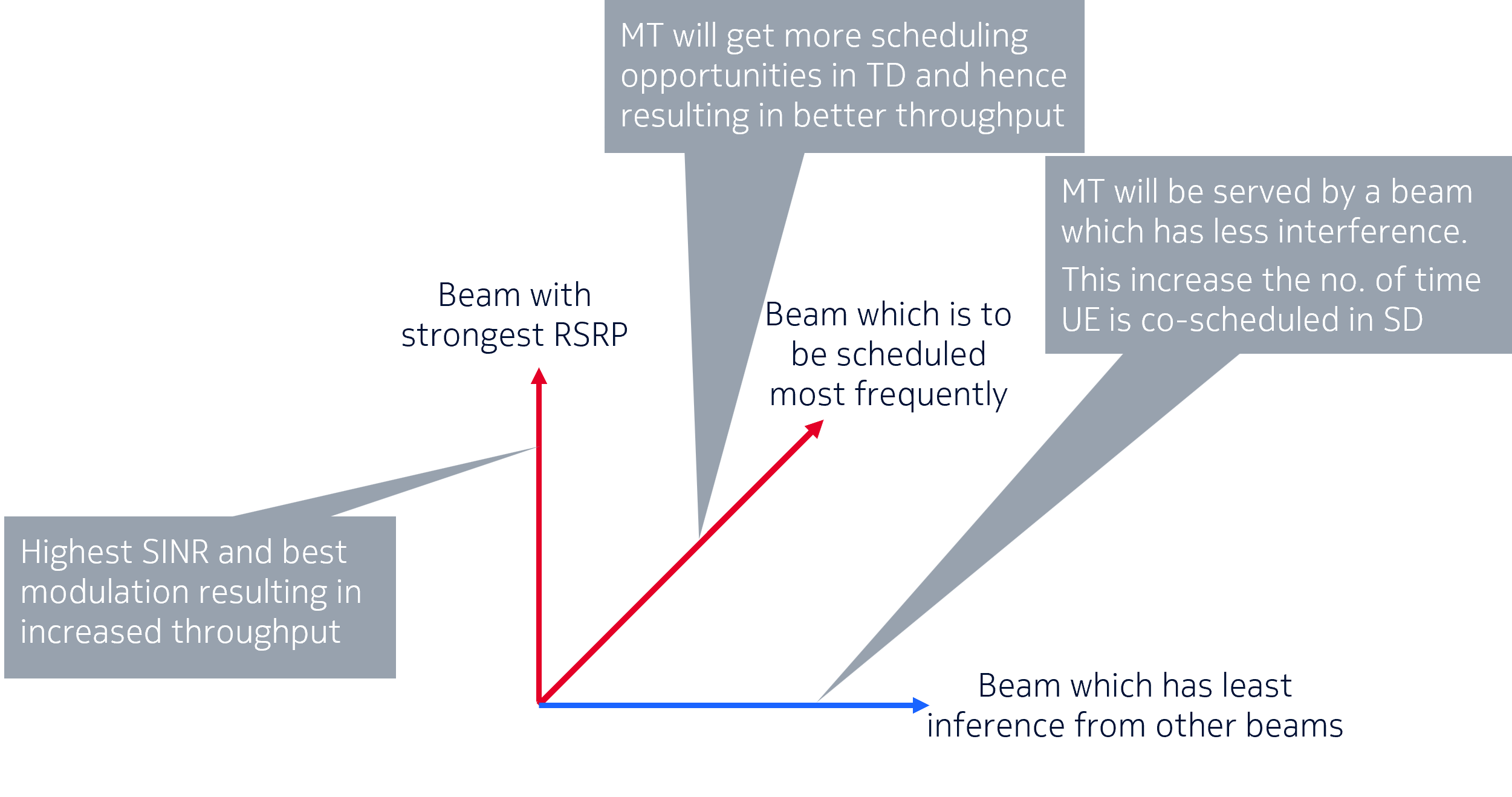}
		\caption{Possible dimensions to be considered in beam-based MU-MIMO beam management procedure apart from traditional RSRP-based legacy approach.}
		\label{fig:dimensions}
    \end{figure}

    The multi-panel configuration at the gNB enhances spatial coverage and diversity but introduces additional complexity into beam selection. Since each panel is constrained by a single RF chain, only one beam per panel can be scheduled per TTI. Consequently, resource blocks (RBs) may be underutilized if the selected beams do not align well with MT traffic demands. As shown in Fig.~\ref{fig:0}, selecting beams solely on the basis of strongest RSRP does not necessarily maximize spectrum efficiency as there are three dimensions shown in Fig. \ref{fig:dimensions}. In certain cases, beams with slightly weaker RSRP but higher scheduling frequency can yield better utilization of available bandwidth. In addition, selecting beams without any expert knowledge on the cross-correlation values does not result in optimized throughput. This observation reveals a key trade-off: maximizing instantaneous link quality by selecting beams with stronger RSRP versus improving long-term efficiency by exploiting beams with higher activation likelihood or lower cross-correlation values. In multi-panel deployments, this trade-off must be carefully optimized, as beam selection decisions across panels are inter-dependent. Hence, advanced algorithms are required to jointly incorporate signal strength, historical activation statistics, and inter-panel beam dynamics to achieve system-level performance gains.

    \setlength{\textfloatsep}{10pt}% Remove \textfloatsep

    As shown in Fig. \ref{fig:dimensions} there are certain dimensions added to the beam selection criteria in beam-based MU-MIMO use-case. Optimizing the system for better throughput and link quality, e.g., SINR could involve clustering MTs onto a single beam via FD multiplexing, which helps mitigate interference by ensuring less frequency resource interference between MTs. Besides, using a popular beam which gets more scheduling opportunities could avoid unwanted delays in buffered packets (since activation of a new beam could result in delays). Additionally, separating the beams efficiently allows for SD multiplexing, and enables two MTs to be co-scheduled using distinct beams. This approach leverages the spatial separation to reduce cross-interference, consequently maximizes throughput/SINR.
    \setlength{\textfloatsep}{10pt}% Remove \textfloatsep

Let us denote the cross-correlation values among beams within different panels by $\rho_{b,j}, b\neq j$ for beam $b$, and $j$. It is known that $\rho_{b,j}$ affects the quality of transmission and eventually achieved throughput significantly. To statistically compute this value, first, we determine the set of beams previously allocated to MTs during past scheduling intervals, i.e., past TTIs. For each beam, the historical activation count is tracked per MT. Any beam that has been selected at least once is classified as a potential interfering beam and included in the interference candidate set. Consequently, the spatial cross-correlation among beams which originates from different antenna panels is computed via the procedure outlined below:
\begin{itemize}
    \item For the first MT (with highest scheduling priority) for which the beam is predicted cross-correlation part of the vector is all zero.
    \item For each MT after the first MT:
    \begin{itemize}
        \item Find the beams which are already assigned to MT for which the beam selection is performed. 
        \item For all the candidate beams considered for the current MT, calculate the sum cross-correlation with already assigned MTs. 
        % \item Normalize the values to be on the interval $[0,1]$. 
    \end{itemize}
\end{itemize}
	
% \subsection{Problem Formulation}
\label{ProbFormulationSec}
In enhanced mobile broadband (eMBB) scenarios, an important design target is to maximize the amount of data successfully delivered to the MT. We formulate this as a per-user maximization under the assumption that the scheduler manages user orthogonality. This requirement can be expressed as the following optimization problem:
\begin{flalign} 
    \textbf{P1: } &
    \max_{b} \; \Gamma_u^{\mathrm{tot}} = \sum_{t=1}^{T} \gamma_t^u(h_b,z'_b,\rho_{b,j})  
    \\ 
    \text{subject to} \quad 
    & \text{C$_1$: } -140 \leq z_b' \leq -44, \quad \forall b \in \mathcal{B}, \nonumber  \\
    & \text{C$_2$: } 0 \leq h_b \leq \frac{T_{\mathrm{bs}}}{T_{\mathrm{TTI}}}, \quad \forall b \in \mathcal{B}, \nonumber \\ 
    & \text{C$_3$: } \rho_{b,j} \in [0,1], \quad \forall b,j \in \mathcal{B}, \; b \neq j, \nonumber 
\end{flalign} 
where $T$ is the total number of TTIs in one beam switching interval $T_{\mathrm{bs}}$, $\Gamma_u^{\mathrm{tot}}$ represents the total throughput of user $u$ during one beam-switching interval, and $\gamma_t^u(h_b,z'_b,\rho_{b,j})$ is the achieved throughput in the $t$-th TTI denoted by $T_{\mathrm{TTI}}$ which is affected by many factors such as the wireless channel gain, interference power, and the gNB's scheduler algorithm \cite{kela2024towards}. Constraint C$_1$ restricts the raw RSRP measurement $z_b'$ to the reporting range specified by 3GPP. Constraint C$_2$ limits the activation counter $h_b$, which records how many times beam $b$ was selected in the interval, with an upper bound equal to the number of TTIs, i.e., $T_{\mathrm{bs}}/T_{\mathrm{TTI}}$ over beam switching interval $T_{\mathrm{bs}}$. Constraint C$_3$ is the cross-correlation coefficients' bound $\rho_{b,j}$ between beams $b$ and $j$ as the probability of co-scheduling. 

Problem \textbf{P1} shows that effective beam selection requires balancing three factors: strong signal strength ($z_b$), the past beam usage ($h_b$), and cross-correlation of beams ($\rho_{b,j}$). Since the search space grows rapidly with the number of beams $B$, direct optimization is computationally highly sophisticated and infeasible in practice. RL offers a scalable alternative, where the state features $\{h_b, z_b, \rho_{b,j}\}$ are used to train an agent that learns a selection policy. Once trained, the decision is obtained by policy inference, avoiding the need for complex iterative optimization.

\vspace{-2mm}
\section{DRL-based Formulation}
\label{DRLSec}
    In RL-aided beam management, the agent seeks to determine an optimal policy $\pi^*:\mathcal{S}\rightarrow\mathcal{A}$ that maximizes the expected communication performance by mapping observed channel states to beamforming actions. The interaction between the agent and the wireless environment is modeled as a Markov Decision Process (MDP), defined as $(\mathcal{S}, \mathcal{A}, \mathcal{R}, \mathcal{P})$, where:
    \begin{itemize}
        \item $\mathcal{S}$ represents the state space, which may include CSI coefficients, RSRP values, or previous beam indices,
        \item $\mathcal{A}$ denotes the discrete action space, corresponding to the set of available beam indices,
        \item $\mathcal{R}:\mathcal{S}\times\mathcal{A}\rightarrow\mathbb{R}$ defines the reward function, which can be designed as the instantaneous spectral efficiency, signal-to-interference-plus-noise ratio (SINR), or achievable throughput,
        \item $\mathcal{P}(s'|s,a)$ specifies the transition probability of reaching state $s'\in\mathcal{S}$ from state $s\in\mathcal{S}$ when action $a\in\mathcal{A}$ is executed.
    \end{itemize}
    The agent’s objective is to maximize the expected discounted cumulative reward $J(\pi) = \mathbb{E}_{\pi}\left[\sum_{t=0}^{\infty} \gamma^t r_t \right]$ where $\gamma \in [0,1)$ is the discount factor, and $r_t$ denotes the immediate communication reward (e.g., spectral efficiency) at time step $t$. The action-value function under policy $\pi$ is
    \begin{equation}
        Q^{\pi}(s,a) = \mathbb{E}_{\pi}\left[\sum_{t=0}^{\infty} \gamma^t r_t \,\bigg|\, s_0 = s, a_0 = a\right],
    \end{equation}
    which satisfies the Bellman expectation equation:
    \begin{equation}
        Q^{\pi}(s,a) = \mathcal{R}(s,a) + \gamma \sum_{s'\in\mathcal{S}} \mathcal{P}(s'|s,a)\, \mathbb{E}_{a'\sim\pi(\cdot|s')}[Q^{\pi}(s',a')].
    \end{equation}
    The optimal beam selection strategy is derived from the optimal action-value function $Q^*(s,a)$, which satisfies the Bellman optimality equation:
    \begin{equation}
        Q^*(s,a) = \mathcal{R}(s,a) + \gamma \sum_{s'\in\mathcal{S}} \mathcal{P}(s'|s,a) \max_{a'\in\mathcal{A}} Q^*(s',a').
    \end{equation}

    In practical mmWave and THz communication scenarios, accurately modeling the transition dynamics $\mathcal{P}(s'|s,a)$ is infeasible due to channel sparsity, user mobility, and hardware impairments. To overcome this limitation, model-free RL algorithms are employed, which directly approximate $Q^*(s,a)$ or its policy counterpart without requiring explicit knowledge of $\mathcal{P}$. This formulation enables adaptive beam selection, where the agent progressively refines its beamforming decisions based on observed RSRP measurements, and gNB-specific information corresponding performance feedback.

    Assuming $K$ candidate beams are considered for RL phase, the raw RSRP measurements are $\mathbf{z}' = [z_1', z_2', \dots, z_B']$ where $z_b'$ corresponds to the measured RSRP of the $k$-th beam. Following the 3GPP specification, these measurements are clipped within the range $[-140,-44]$ dBm and then normalized as
    \begin{flalign}
        \mathbf{z} = \frac{\mathbf{z}' + 140}{96}, \quad 
        \mathbf{z} = [z_1, z_2, \dots, z_B].
    \end{flalign}
    In addition, the beam activation history is represented as
    \begin{flalign}
        \mathbf{h} = [h_1,h_2,\dots,h_B],
    \end{flalign}
    where $h_b$ denotes the normalized usage counter of the $k$-th beam within the most recent beam-switching interval. Each interval spans approximately $40$ ms, which contain multiple TTIs. During each TTI, the scheduler selects an active beam, and the cumulative usage statistics over TTIs are stored in $\mathbf{h}$.
    
    The beam selection task is addressed within an RL framework by defining the state, action, and reward functions. We investigate them in detail as follows:
    
    \subsubsection{States}
    The RL agent constructs its decision-making context by forming a state representation that integrates multiple dimensions of information. Similar to our previous work \cite{Hashemi2024}, the state vector incorporates:  
    (i) normalized RSRP values of candidate beams,  
    (ii) beam activation frequencies over the preceding switching interval, and  
    (iii) pairwise cross-correlation indicators between serving beams of active MTs, which reflect their likelihood of being scheduled jointly.  
    
    Thus, the state at time step $t$ is
    \begin{equation}
        \mathbf{s}_t = \big\{h_b, \, z_b, \, \rho_{b,j} \; \big| \; \forall b,j \in \{1,2,\dots,B\}, \; b \neq j \big\},
    \end{equation}
    where $z_b$ denotes the normalized RSRP of beam $b$, $h_b$ represents the normalized activation frequency of beam $b$, and $\rho_{b,j}$ captures the cross-correlation between beams $b$ and $j$ within different antenna panels. The rationale for including activation statistics, rather than buffer-related traffic descriptors, is that past beam usage patterns provide a more reliable indicator of future scheduling tendencies. The overall dimensionality of the state space grows with both the number of beams $B$ and the cross-correlation terms.

    \subsubsection{Actions}
    At each decision epoch, the agent selects one beam from the set of available candidates. The policy network outputs action-values over the $B$ beam indices, and the action is obtained by selecting the argmax of the output. The action $a_t$ at the time step $t$ is defined as $a_t \in \{1,2,\dots,B\}$ where $a_t = b$ indicates the $b$-th beam is activated during the slot $t$.

    \subsubsection{Reward}
    Let $B_k^t$ denote the number of data bytes successfully delivered to the user $k$ at the time step $t$. The reward function is based on the observed throughput of the MT. In addition, to stabilize the training process, the reward is normalized with respect to the maximum data bytes delivered across all users. The reward assigned to the agent is
    \begin{equation}
        r_t = \frac{B_k^t}{\max_j (B_j^t)},
    \end{equation}
    where $\max_j (B_j^t)$ denotes the peak transmission outcome among all active users at time $t$. This normalization prevents the reward magnitudes from becoming unbounded and accelerates the convergence of the learning algorithm \cite{sutton1998reinforcement}.
    
    The optimal beam selection policy within the beam management framework in beam-based MU-MIMO is obtained using the double deep Q-Network (DDQN) method to approximate the action-value function and identify the beam that maximizes the expected Q-value \cite{sutton1998reinforcement}. The state representation utilized by the agent comprises normalized RSRP measurements, a beam activation history, and cross-correlation values as described. 
    % However, applying full-scale RL methods to scenarios with a large number of beams presents several challenges. Notably, tuning the hyperparameters of a DDQN becomes increasingly difficult as the size of the action space grows, requiring extensive system-level simulations to ensure stable convergence. Consequently, lightweight or simplified RL approaches that demand fewer hyperparameter adjustments are especially valuable in environments with large beam sets, where the action space is correspondingly extensive.

\vspace{-1.5mm}
\section{Numerical Results}
\label{NumericalSec}
    In this section, we present the system level simulation results. We use a realistic system simulation platform to test the proposed beam selection approach\footnote{For details on used simulation engine please refer to \cite{Abinader2021,Pedersen2024}.}. The objective of the agent is to maximize the total throughput in DL with the FTP3 traffic model where the offered load per MT is 21 Mbps.
    % \textcolor{red}{I assume that 15 Mbps in the parameter table is the offered load per UE? What is the packet size?}. 
    We assume that the model is on the gNB side unless otherwise stated. 
    % To benefit from \textcolor{red}{"recent advances" in RL/ML mentioned 5 times already. Maybe we could reduce such phrase repetitions. Also, IMO no need to re-explain PPO again here.} recent advances in RL algorithms, we have developed and employed recurrent PPO that are on-policy methods that adapt themselves according to obtained experience immediately (in a short time) unlike conventional off-policy methods, e.g., DDQN.  
    A macro-cellular scenario composed of a standard hexagonal grid of three-sector 5G NR gNBs with an inter-site distance of 200 meters is assumed along with a carrier frequency of 30 GHz as in 3GPP guidelines for FR2. Table \ref{tab1} summarized the important parameters assumed for the simulations. To enable beam-based MU-MIMO over multiple antenna panels, we leveraged hybrid beamforming framework by pairing the MTs based on a threshold for cross-correlation with the aim of maximizing the proportional fairness metric in the time/frequency domain. In this method, the MTs are selected to be paired with the first MT if they meet a correlation threshold condition. The role of RL-based beam selection is to manipulate the beams associated with each MT before performing hybrid beamforming.

    \begin{table}[t]
        \caption{Simulation configuration.}
        \centering
        \begin{tabular}{ l  m{4cm} }
            \hline
            Parameter & Value \\ \hline
            Deployment scenario & 3GPP Dense Urban Macro \cite{3gpp_tr_38913}, \\ 
                                & 3D channel model in \cite{3GPPTR38.901} \\
            gNB transmit power & $40$ dBm  \\
            Number of MTs ($U$) & $210$  \\ 
            Number of gNBs ($M$) & $21$ (7 sites, 3 sectors/site) \\
            Antenna panels per gNB ($M_p$) & $3$ panels per sector \\
            Beams per panel ($L_p$) & $16$ beams  \\
            Total number of beams ($B$) & $B=L_p\times M_p=48$ \\
            % gNB antenna $(M, N, P, Mg, Ng)$ & $(8, 8, 2, 1, 3)$ \\
            Beam-switching interval ($T_{\mathrm{int}}$) & $40$ ms \\
            Hybrid beamforming algorithm &  Correlation threshold \\
            MT pairing correlation threshold & 0.4 \\
            MT velocity & $3$ km/h \\
            Carrier frequency & FR2, $30$ GHz \\ 
            System bandwidth ($W$) & $200$ MHz \\ 
            Subcarrier spacing & $60$ KHz \\ 
            Receiver noise figure & $3$ dB \\
            Thermal noise density ($N_0$) & $-174$ dBm/Hz \\ 
            Traffic model & FTP3, $21$ Mbps target rate \\
            Packet size & $600$ Bytes \\
            Training batch size & $32$ samples \\
            Experience replay size & $5000$ \\
            Learning rate (Adam optimizer) & $10^{-4}$  \\
            Activation function &  $ReLU(\cdot)$ \\
            Neural network architecture & $(\text{input},128,256,\text{output})$ \\
            Training epoch & 4 \\
            \hline
        \end{tabular}
        \label{tab1}
    \end{table}
\setlength{\textfloatsep}{10pt}% Remove \textfloatsep

\begin{figure}[t]
    \centering
    \includegraphics[width=\linewidth]{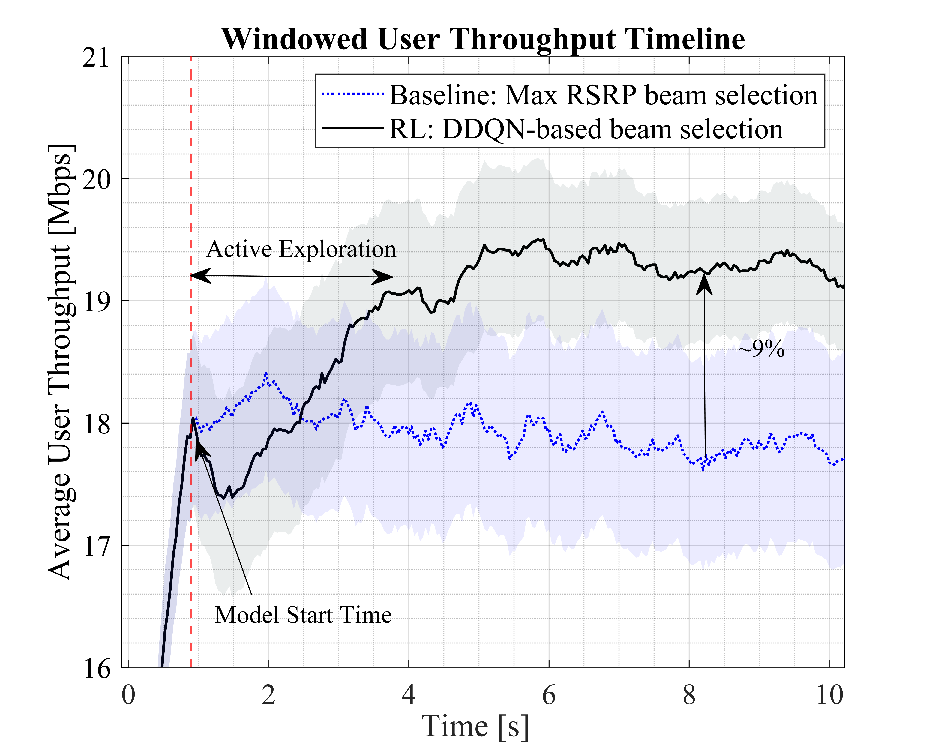}
    \caption{Impact of exploration during training on the average MT throughput (total no. of MTs in network $U=210$).}
    \label{fig:ExplorationEffect}
\end{figure}

Exploratory beam selection during training can impact performance, potentially causing it to temporarily fall below baseline levels, as shown in Fig. \ref{fig:ExplorationEffect}. Due to the closed-loop nature of RL training, exploration is needed to learn the relationship between selected beams and the observed reward, i.e., throughput. Fig. \ref{fig:ExplorationEffect} shows the impact of exploring various beams on the average MT throughput over time (there are 210 MTs in the network). During active exploration, the average throughput is approximately 10\% below the model's converged average throughput and falls below the baseline method.  
\setlength{\textfloatsep}{10pt}% Remove \textfloatsep

\begin{figure}[t]
    \centering
    \includegraphics[width=\linewidth]{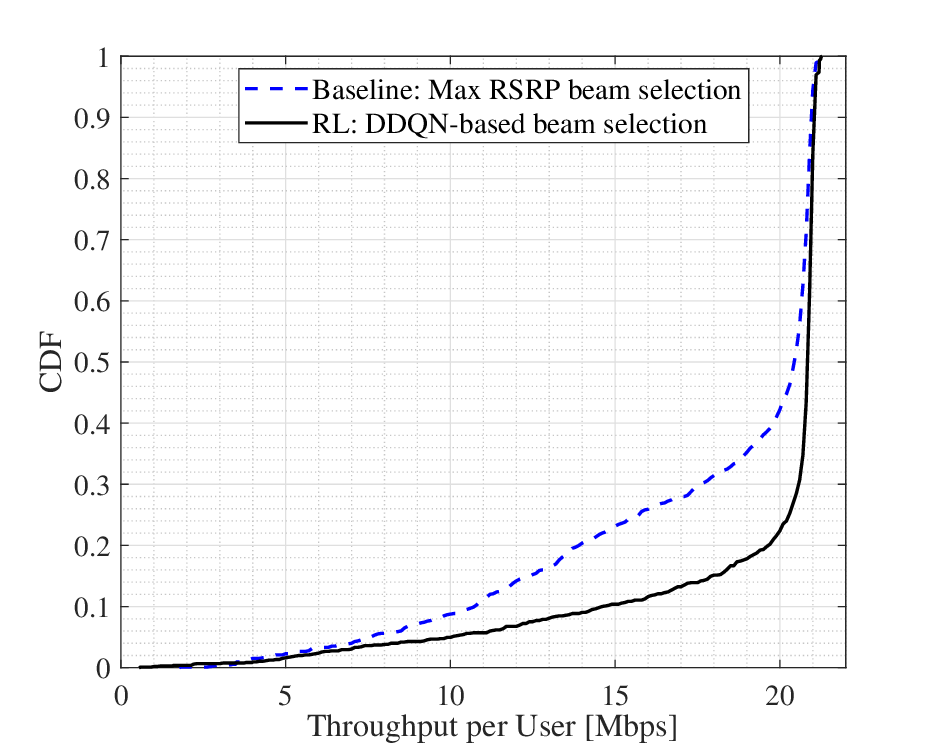}
    \caption{The CDF of user throughput ($U=210$).}
    \label{fig:CDF}
\end{figure}

Fig. \ref{fig:CDF} presents the cumulative distribution function (CDF) of user throughput in a network comprising 210 MTs. The figure compares two beam selection strategies: the baseline method, which utilizes maximum RSRP for beam selection, and a RL approach employing a DDQN for beam selection. The CDF curves illustrate that the RL-based DDQN beam selection consistently outperforms the baseline method, achieving higher throughput for users across the distribution. Specifically, the RL approach results in a steeper CDF curve, indicating that a larger proportion of users experience higher throughput compared to the baseline method. This demonstrates the effectiveness of the RL-based approach in optimizing network performance and enhancing user experience.
\setlength{\textfloatsep}{10pt}% Remove \textfloatsep

\begin{figure*}[t]
    \centering
    \includegraphics[width=\linewidth]{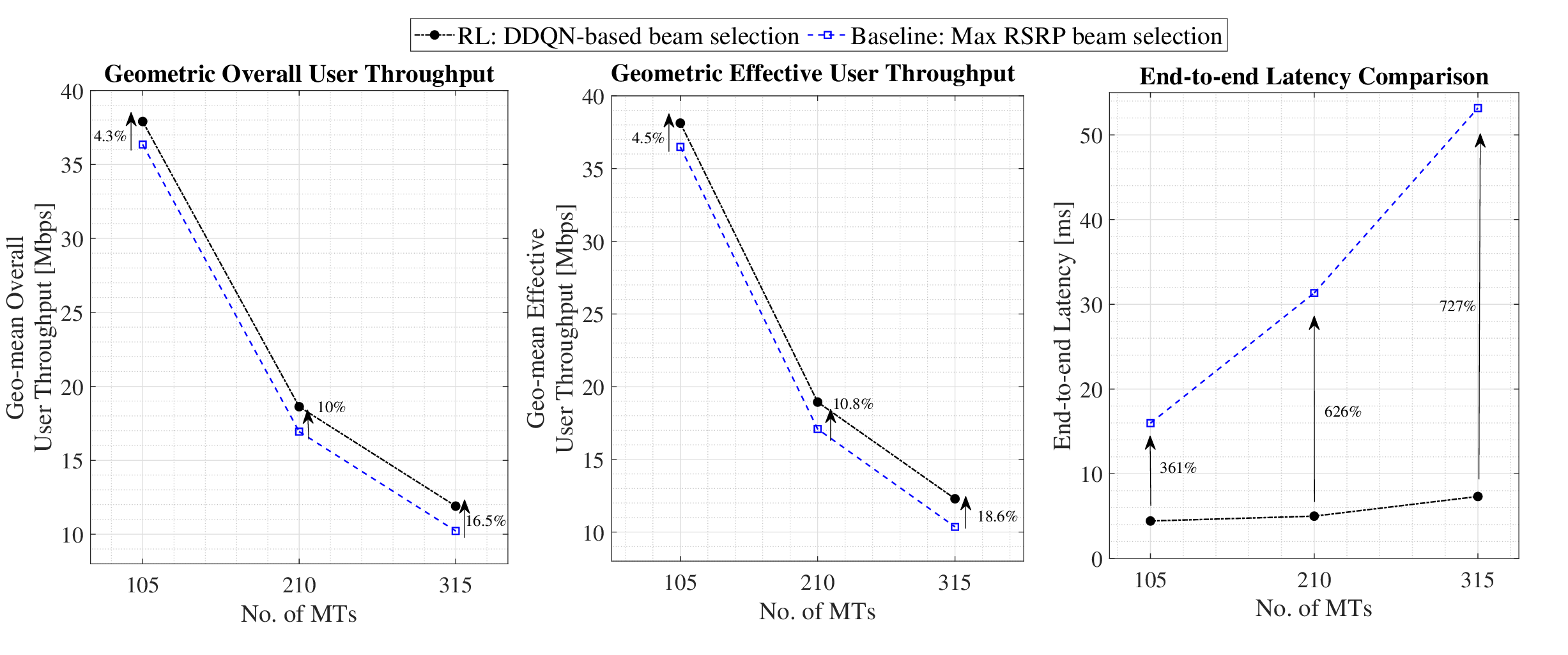}
    \caption{Comparison between geometric mean user throughput, and the end-to-end latency with RL-DDQN vs. baseline.}
    \label{fig:overall}
\end{figure*}

In Fig. \ref{fig:overall} a comparison is made between the RL-based beam selection and the strongest RSRP beam selection method, i.e., the baseline (legacy approach). As shown, RL-based framework provides higher overall/effective geometric mean user throughput. The effective throughput is calculated as the ratio of the total amount of data sent to the MT vs. the total time the MT was scheduled for reception. In addition, overall throughput is calculated as the ratio of the total amount of data sent to the MT vs. the total time the MT was active. To show that all MTs have achieved optimized throughput, we compare geometric mean user throughput w.r.t. baseline beam selection. The geometric mean (GM) is defined as
\begin{equation}
    \text{GM} = \sqrt[U]{\Gamma^{\text{tot}}_1 \Gamma^{\text{tot}}_2 \cdots \Gamma^{\text{tot}}_U}, \quad \Gamma^{\text{tot}}_u\neq 0 \quad \forall u \in \mathcal{U}
\end{equation}
where $\Gamma^{\text{tot}}_u$ is the computed throughput value for MT$_u$. In addition, because of optimized beam selection and improved scheduling opportunities for MTs with RL, end-to-end latency has been significantly reduced. More specifically, RL-based beam selection optimization provides up to 5-16\% throughput gain and 3-7x improvement in end-to-end latency compared to baseline, i.e., maximum RSRP beam selection method. 
%The main reason for such latency enhancement is under the learned policy of the RL agent which tries to schedule, and groups MTs with sub-optimal beams (in terms of the RSRP) so that the gNB allocates resources immediately without buffering the users' packets for a long time to activate certain beams. 
The main reason for this improvement in latency is the learned policy of the RL agent. This agent schedules and groups MTs with sub-optimal beams, in terms of RSRP, allowing the gNB to allocate resources immediately. As a result, users' packets do not need to be buffered for extended periods while waiting for certain beams to activate.

\begin{figure}[t]
    \centering
    \includegraphics[width=\linewidth]{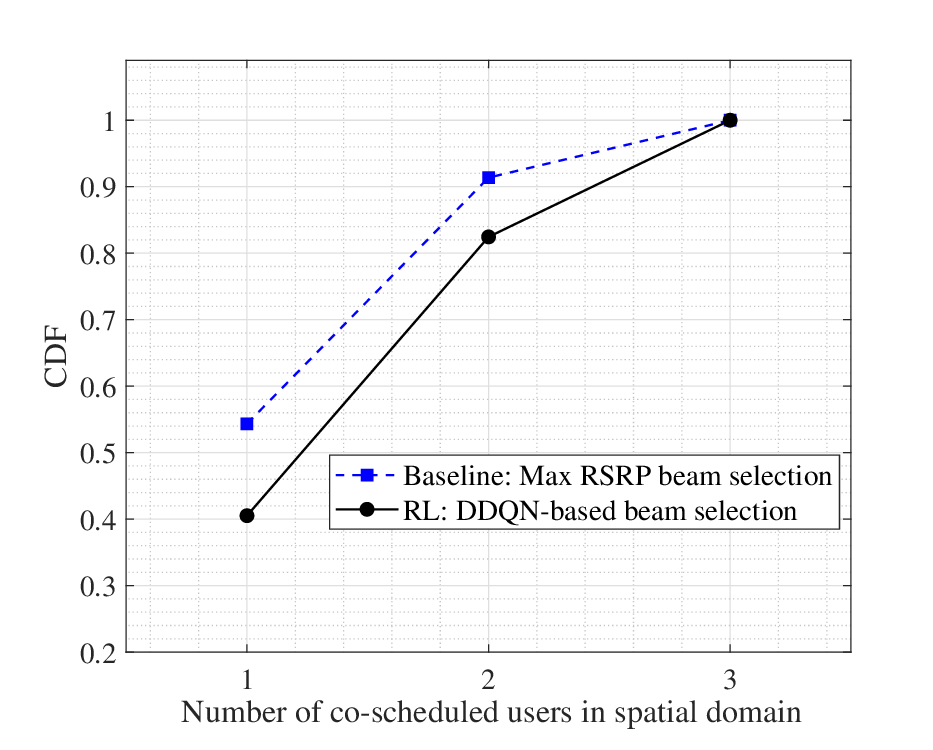}
    \caption{The CDF of co-scheduled users in spatial domain.}
    \label{fig:SD}
\end{figure}

The RL agent learns this policy by leveraging the spatial, temporal, and received power characteristics of the beams. For example, as shown in Fig. \ref{fig:SD}, the RL agent has grouped more MTs than the baseline, which improves the scheduling opportunities. Note that the maximum number of co-scheduled users is always limited by the antenna panels in beam-based MU-MIMO, i.e., $M_p=3$.

\setlength{\textfloatsep}{10pt}% Remove \textfloatsep

\setlength{\textfloatsep}{10pt}% Remove \textfloatsep

 	\vspace{-3mm}
   \section{Conclusion}
   \label{ConslusionSec}
   In this study, we have presented a DRL-based framework for beam management in multi-panel mmWave Radio Access Networks aimed at improving user throughput and minimizing latency. By employing a scalable RL approach, we have addressed the complexities involved in optimized beam selection within MU-MIMO systems as defined in \textbf{P1}, where traditional optimization methods are infeasible. Our proposed solution demonstrated considerable gains in throughput, achieving up to a 16\% increase, and substantial reductions in end-to-end latency by factors of 3-7x compared to baseline beam selection techniques. These findings validate the effectiveness of using RL in dynamic wireless environments and offer promising insights into the management of complex network scenarios to enhance the user experience on mmWave frequencies. Future research can explore further optimizations and the integration of additional dimensions of channel state information or involve digital beamforming techniques.

    % \appendices
    
    % \section{Proof of}
    % \label{appA}
	%	\newpage
    \vspace{-2.5mm}
    \bibliographystyle{IEEEtran}% such as plain
% 	\begin{spacing}{1.33} %% COMPRESSION TWEAK
    \bibliography{refs}
    % \bibliography{references3}
    % \end{spacing}
	
\end{document}